\documentclass[preprint,pra,showpacs,nofootinbib]{revtex4}
\usepackage{mathrsfs}
\usepackage{float,epsfig}
\usepackage{dcolumn}% Align table columns on decimal point
\usepackage{bm}% bold math
\usepackage{graphicx}% Include figure files
\usepackage{dcolumn}% Align table columns on decimal point
\usepackage{bm}% bold math
\usepackage{amsmath,amssymb,amsthm}
\begin{document}
\title{\bf Casimir-Polder-like force on an atom  outside a Schwarzschild black hole }
\author{  Jialin Zhang$^{1}$ and Hongwei Yu$^{1,2}$
%\footnote{Corresponding author}
}
\affiliation{$^1$ Institute of Physics and Key Laboratory of Low
Dimensional Quantum Structures and Quantum
Control of Ministry of Education,\\
Hunan Normal University, Changsha, Hunan 410081, China \\
$^2$ Center for Nonlinear Science and Department of Physics, Ningbo
University,  Ningbo, Zhejiang 315211, China}

%\date{\today}

\begin{abstract}

We calculate, in the framework of open quantum systems, the  ground
state energy-level shift  for a static two-level atom outside a
spherically symmetric black hole in interaction with fluctuating
massless scalar fields in the Boulware and Unruh vacuums. We find
that the energy-level shift is position dependent and thus gives
rise to a force on the atom besides the classical gravitational
force. For the case of the Boulware vacuum that represents  a star
which has not collapsed through its event horizon, this force  is
attractive  near the horizon and is repulsive far away from the
black hole with a  behavior of $r^{-3}$.   For the case of the
Unruh vacuum which represents a radiating black hole, we find that
the contribution to the Casimir-Polder-like force due to the
presence of Hawking radiation is always attractive and,
remarkably, this attractive force diverges at the event horizon.

\end{abstract}
 \maketitle

\section{Introduction}
%%%%%%%%%%%%%%%%%%%%%%%%%
 The shift of the energy-levels of an atom which is caused by the coupling of the atom with  the quantum vacuum is
 one of the striking manifestations of the existence of zero-point fluctuations, and it is known that this
shift can be modified by the presence of cavities~\cite{Meschede90}
and the non-inertial motion of the atom
itself~\cite{Audretsch95,Passante98,L.Rizzuto07,ZhuYu10}. One of the
observable examples of such shifts is the Casimir-Polder force
between a neutral electric polarizable atom and a conducting plate,
which is a result of position-dependent energy-level shift caused by
the modification of vacuum fluctuations that arise because of
reflection of vacuum field modes at the conducting boundaries.

So far, the Casimir-Polder force, which is a consequence  of the
position-dependent level shifts of an atom, is regarded as a result
of the reshaping of vacuum fluctuations induced by the reflection of
field modes at boundaries in flat spacetimes. Since the vacuum field
modes are also scattered by the curvature in a curved spacetime,
such as in the background of a black hole,  a question naturally
arises as to whether  the scattering of vacuum field modes off the
curvature will also produce a Casimir-Polder-like force on an atom
outside a back hole, and this is exactly what we are going to
address in the present paper.

Our calculation of the energy-level shifts is based upon the
framework of open quantum systems\cite{BP}, where a static  two-level
atom outside a Schwarzschild back hole is treated as an open quantum
system in interaction with the external reservoir of fluctuating
vacuum massless scalar fields. As for any open system, the full
dynamics of the atom can be obtained from the complete time
evolution describing the total system (atom plus a reservoir of
external vacuum fields) by integrating over the field degrees of
freedom, which are in fact not observed. It is worth noting here that the
open quantum system approach has also been applied to studying other
quantum effects in curved spacetime, such as the Hawking radiation
of a black hole~\cite{zhjl} and the Gibbons-Hawking effect of de
Sitter spacetime~\cite{yu4}.

When a curved space-time is considered as opposed to a flat one, a
delicate issue  arises as to how the vacuum state of the quantum
fields is specified. In this paper, we  deal with two vacuum states
of  the scalar fields; namely, the Boulware vacuum and  the Unruh
vacuum. Let us note that the Boulware vacuum  would be the vacuum
state outside a massive spherical body of radius only slightly
larger than its Schwarzschild radius and the Unruh state that
best approximates the vacuum following the gravitational collapse of
a massive body to a black hole. So, we are going to compute the
force on a static atom as a result of the modified vacuum
fluctuations both outside a star which has not collapsed
through its event horizon and a black hole.

The paper is organized as follows: In the next section, we  will
review the basic formalism of open quantum systems, the derivation
of the master equation describing the system of the atom plus
external vacuum scalar fields in the weak-coupling limit, and the
reduced dynamics it generates for the finite-time evolution of the
atom. In Sec. III, we calculate the  radiative energy shift of the
ground state and the resulting force on the atom. Finally, we
conclude in Section IV.

%%%%%%%%%%%%%%%%%%%%%%%%%%%%%
\section{the basic formalism}

We  consider the evolution in the proper time of a static two-level
atom interacting with vacuum massless scalar fields outside a
Schwarzschild black hole and assume the combined system (atom +
external vacuum fields) to be initially prepared in a factorized
state, with the atom held static in the exterior region of the black
hole and the fields in the Boulware or Unruh vacuum state. When the quantum
system of a two-level atom interacts weakly with the environment,
the reduced dynamics can be
 obtained by eliminating the environment degrees of freedom, yielding  an evolution equation of the atom that satisfies the
master equation\cite{Benatti0,Benatti1,Benatti2}.  We take the total
Hamiltonian for the complete system to have the form
\begin{equation}
\label{H}
 H=H_S+H_\phi+ H_I\; ,
 \end{equation}
where  $H_S$ denotes the free Hamiltonian of the atom and  $H_\phi$
that of the environment(a bath of fluctuating quantum fields). In
fact,  $H_\phi$ can be chosen as the standard Hamiltonian of
massless, free scalar fields, the details of which are not relevant
here. In order to make our discussion generic, we will postpone the
specification of $H_S$ until later and suppose that the interaction
Hamiltonian has the general form
\begin{equation}\label{HIG}
H_I=\mu\sum_\alpha A_\alpha\otimes{B}_\alpha\;,
\end{equation}
 where $A_\alpha$ and
$B_\alpha$ represent respectively the dynamical variables of the
atom and of the environment. It should be pointed out that the
coupling constant $\mu$ should be small, and this is required by our
assumption that the interaction of the atom with the scalar fields
is weak.

Initially,  the complete system is described by the total density
 $\rho_{tot}=\rho(0)\otimes\rho_B\;,$
where $\rho(0)$ is the initial reduced density matrix of the atom,
and $\rho_B$ is the state of the environment.
  In the frame of the atom,  the evolution in the proper time $\tau $ of the total density
  matrix
$\rho_{tot}$ of the complete system satisfies
\begin{equation}
\frac{\partial\rho_{tot}(\tau)}{\partial\tau}=-iL_H[\rho_{tot}(\tau)]\;,
\end{equation}
where the symbol $L_H$ represents the Liouville operator associated
with $H$
\begin{equation}
L_H[S]\equiv[H,S]\;.
\end{equation}
The dynamics of the atom can be obtained by tracing over the field
degrees of freedom,  that is, by applying the trace projection to the
total density matrix $\rho(\tau)={\rm
{Tr}}_{\Phi}[\rho_{tot}(\tau)]\;$.

In the limit of weak-coupling which we assume in the present paper,
the reduced density  is found to obey the master equation in the
interaction picture~\cite{BP,Lindblad}
%%%%%%%%%%%%%%%%%%%%%
\begin{equation}\label{master}
\frac{d}{d\tau}\rho_{S}(\tau)=-i[H_{LS},\rho_{S}(\tau)]+{\cal
{D}}(\rho_{S}(\tau))\;,
\end{equation}
where $H_{LS}$ provides a Hamiltonian contribution to the dynamics
of the system and  is the so-called  Lamb shift Hamiltonian,
whereas, ${\cal {D}}(\rho_{S}(\tau))$ is called the dissipator of the
master equation. Let us note that here $H_{LS}$ is given by
\begin{equation}\label{HLS}
H_{LS}=\mu^2\sum_\omega\sum_{\alpha,\beta}S_{\alpha\beta}(\omega)A^+_\alpha(\omega){A_\beta(\omega)}
\end{equation}
and the system operator can be decomposed as \cite{BP}
\begin{equation}
A_\alpha=\sum_{\omega}A_\alpha(\omega)=\sum_{\omega}A^+_\alpha(\omega)\;,
\end{equation}
with
\begin{equation}\label{Aw}
A_\alpha(\omega)=\sum_{\nu\;'-\nu=\omega}\Pi(\nu)A_\alpha\Pi(\nu\;')\;,
\end{equation}
where the  operator $\Pi(\nu)$ denotes  the projection onto the
eigenspace  belonging to the eigenvalue $\nu$ of $H_S$.  It is easy
to see that
\begin{equation}
[H_S,A^+_\alpha(\omega)A_\beta(\omega)]=0\,,
\end{equation}
that is, $H_{LS}$ has the same eigenstates as $H_S$. Meanwhile, the
function $S_{\alpha\beta}(\omega)$ can be written as~\cite{BP}
\begin{equation}
S_{\alpha\beta}(\omega)= \frac{i}{2}
{\cal{G}}_{\alpha\beta}(\omega)-i\Gamma_{\alpha\beta}(\omega)\;,
\end{equation}
where ${\cal{G}}_{\alpha\beta}(\omega)$ is the Fourier transform of
the reservoir correlation function
(${\langle}B^+_\alpha(s)B_\beta(0)\rangle$)
\begin{equation}\label{gamma1}
{\cal{G}}_{\alpha\beta}(\omega)=\int^\infty_{-\infty}dse^{i\omega{s}}{\langle}B^+_\alpha(s)B_\beta(0)\rangle\;
\end{equation} and $\Gamma_{\alpha\beta}(\omega)$ denotes the one-side Fourier transform
\begin{equation}
\Gamma_{\alpha\beta}(\omega)=\int^\infty_0dse^{i\omega{s}}{\langle}B^+_\alpha(s)B_\beta(0)\rangle\;.
\end{equation}
Then,  with the help of
\begin{equation}
\frac{1}{1\mp{i}\epsilon}=P\frac{1}{x}\pm{i}\pi\delta(x)\;,
\end{equation}
where P denotes the  Cauchy principal value,  it is easy to prove
that
\begin{equation}\label{Salphabeta}
S_{\alpha\beta}(\omega)=-\frac{P}{2\pi}\int^\infty_{-\infty}\frac{{\cal{G}}_{\alpha\beta}(z)}{z-\omega}dz\;.
\end{equation}
The dissipator of the master equation takes the form
\begin{equation}
{\cal{D}}(\rho_{S})=\mu^2\sum_\omega\sum_{\alpha,\beta}{\cal{G}}_{\alpha\beta}(\omega)\bigg(A_\beta(\omega)\rho_{S}A^+_\alpha(\omega)
 -\frac{1}{2}\{A^+_\alpha(\omega)A_\beta(\omega),\rho_{S}\}\bigg)\;.
 \end{equation}
%%%%%%%%%%%%%%%%%%%%%%%%%%%%%%%%%%%%%%%%%
 The Hamiltonian $H_S$ of a two-level
atom can be generically written as~\cite{Benatti1}
 \begin{equation}
 H_S={{\omega_0}\over 2}\sum_{i=1}^3n_i\sigma_i\;,
 \end{equation}
where $\sigma_i\; (i=1,2,3)$ are the Pauli matrices, $\omega_0$ is the
energy-level spacing and $\mathbf{n} =(n_1,n_2,n_3)$ is a unit vector.
Now, we let $\mathbf{n} =(0,0,1)$ (i.e., $H_S={\omega_0}
\sigma_3/2\;$) for simplicity. The interaction Hamiltonian $H_I$ is
taken as
\begin{equation}\label{HI}
H_I=\mu \sum_{\alpha=0}^{3}\sigma_{\alpha}\Phi_\alpha(x)\;,
\end{equation}
where $\sigma_0$ is the unit matrix and external fields are
represented by $\Phi_\alpha(x)$.  Comparing it  with the general
form Eq.~(\ref{HIG}), we find that $A_\alpha=\sigma_\alpha\;,$
$B_\alpha=\Phi_\alpha(x)\;.$

%%%%%%%%%%%%%%%%%%%%%%%%%%%%%
For the sake of simplicity, we now assume that the field
correlation functions are diagonal such that~\cite{Benatti1}
\begin{equation}
G^+(x-y)\delta_{\alpha\beta}=\langle\Phi_\alpha(x)\Phi_\beta(y)
\rangle\label{twogreen}\;.
\end{equation}
Therefore, its  Fourier  and Hilbert transforms  can be expressed
respectively as
\begin{equation}
{\cal G}(\omega)=\int_{-\infty}^{\infty} d\tau \,
e^{i{\omega}\tau}\, G^{+}\big(x(\tau)\big)=\delta_{\alpha\beta}{\cal
G}_{\alpha\beta}(\omega)\;, \label{fourierG}
\end{equation}
and
\begin{equation}
{\cal K}(\omega)= \frac{P}{\pi i}\int_{-\infty}^{\infty} d\lambda
\frac{ {\cal G}(\lambda) }{\lambda-\omega}
=2i\delta_{\alpha\beta}S_{\alpha\beta}(\omega)\;. \label{kij}
\end{equation}
It should be pointed out that here the repeated index does not mean summation.
%%%%%%%%%%%%%%%%%%%%%%%%%%%%%%%%%%%%%%%%%%%
Letting $|+\rangle$ and $|-\rangle$ denote the excited  and the ground
states, respectively, we have
\begin{equation}
A_\alpha(\omega_0)=A_\alpha^+(-\omega_0)=|-\rangle\langle-|\sigma_\alpha|+\rangle\langle+|\;.
\end{equation}
Since the energy spacing is $\omega_0$, according to Eq.~(\ref{Aw}),
the summation over $\omega$ in Eq.~(\ref{HLS}) just contains two
terms; namely, $\omega=\omega_0$ and $\omega=-\omega_0$. As a
result,  Eq.~(\ref{HLS}) can be rewritten as
\begin{eqnarray}\label{HLAW}
H_{LS}&=&\mu^2\sum_\omega\sum_{\alpha,\beta}S_{\alpha\beta}(\omega)A^+_\alpha(\omega){A_\beta(\omega)}\nonumber
\\&=&-i\frac{\mu^2}{2}{\cal{K}}(\omega_0)\sum_{\alpha=0}^3|+\rangle\Big|\langle+|\sigma_\alpha|-\rangle\Big|^2\langle+|
-i\frac{\mu^2}{2}{\cal{K}}(-\omega_0)\sum_{\alpha=0}^3|-\rangle\Big|\langle-|\sigma_\alpha|+\rangle\Big|^2\langle-|\nonumber\\
&=&-i\mu^2{\cal{K}}(\omega_0)|+\rangle\langle+|-i\mu^2{\cal{K}}(-\omega_0)|-\rangle\langle-|
\end{eqnarray}
This shows that the energy-level shifts of the ground state and
excited state are $\delta E_- =-i\mu^2{\cal K}(-\omega_0)$ and
$\delta E_+=-i\mu^2{\cal K}(\omega_0)$, respectively, which are
caused by the coupling of the atom to the vacuum fluctuation of the
fields. The relative energy shift (the Lamb shift) is then
\begin{equation}\label{14}
\Delta=i\mu^2[{\cal K}(-\omega_0)-{\cal K}(\omega_0)]\,.
\end{equation}

\section{Casimir-Polder-like force for Boulware vacuum }
In this section, we apply the open quantum system formalism
developed in the preceding section to address the issue of
finite-time evolution of a static two-level atom interacting with
vacuum scalar fields outside a spherically symmetric black hole and
calculate the energy-level shift of the ground state. The line
element for a Schwarzschild black hole reads
\begin{equation}
ds^2=g_{\mu\nu}x^{\mu}x^{\nu}=\bigg(1-\frac{2M}{r}\bigg)dt^2-\frac{dr^2}{1-2M/r}-r^2(d\theta^2+\sin^2\theta{d\phi^2})\;.
\end{equation}
Let us start with the Boulware vacuum which is deemed to be the
natural vacuum outside a massive body and reduces to the usual
Minkowski vacuum at infinity. The correlation function for the
scalar field in the Boulware vacuum is~\cite{green1,wightman3}
\begin{equation}
 G^+_B(x,x')=\sum_{lm}\int_{0}^{\infty}\frac{e^{-i\omega \Delta{t}}}{4\pi\omega}|\,Y_{lm}(\theta,\phi)\,|^2\big[|\,\overrightarrow{R_l}(\omega,r)\,|^2
+|\,\overleftarrow{R_l}(\omega,r)\,|^2\big]d\omega\;,
 \end{equation}
and the corresponding Fourier transform with  the proper time $\tau$
of the atom  reads
\begin{eqnarray}
\mathcal{G}_B({\lambda})&=&\int^{\infty}_{-\infty}e^{i\lambda{\tau}}{G_{B}}^+[x(\tau)]d\tau\nonumber\\
&=&\sum_{ml}\int^{\infty}_{0}\frac{d\omega}{2\omega}\delta({\lambda}-\omega/\sqrt{1-2M/r})|\,Y_{lm}(\theta,\phi)\,|^2
\big[|\,\overrightarrow{R_l}(\omega,r)\,|^2
+|\,\overleftarrow{R_l}(\omega,r)\,|^2\big]\;.\label{b-green}
\end{eqnarray}

Using  Eq.~(\ref{kij}), we can calculate  $\delta{E}_-$ for the
Boulware vacuum; for convenience, we denote it by $\delta{E}_-^{B}$.
Since the exact form of functions ${R_l}(\omega,r)$ is not known, a
generic expression for $\delta{E}_-^{B}$ is not possible to obtain.
In what follows, we only examine the behavior of $\delta{E}_-^{B}$
at two asymptotic regions, i.e., close to the black hole horizon and
at infinity. In order to do this, let us recall that\cite{wightman3}
\begin{equation} \label{asymp2}
\sum_{l=0}^\infty\,(2l+1)\,|\overrightarrow{R}_l(\,\omega,r\,)\,|^2\sim\left\{
                    \begin{aligned}
                 &\frac{4\omega^2}{1-\frac{2M}{r}}\;,\;\;\;\quad\quad\quad\quad\quad\quad\quad r\rightarrow2M\;,\cr
                  &\frac{1}{r^2}
\sum_{l=0}^\infty(2l+1)\,|\,{B}_\ell\,(\omega)\,|^2\;,\quad\;r\rightarrow\infty
                  \;,
                          \end{aligned} \right.
\end{equation}

\begin{equation} \label{asymp3}
\sum_{l=0}^\infty\,(2l+1)\,|\overleftarrow{R}_l(\,\omega,r\,)\,|^2\sim\left\{
                    \begin{aligned}
                 &\frac{1}{r^2}\sum_{\ell=0}^\infty(2l+1)\,|\,{B}_\ell\,(\omega)\,|^2,\quad\;&r\rightarrow2M\;,\cr
                  &4\omega^2/(1-2M/r)\sim4\omega^2,\;\;\;\;\quad\quad\quad &r\rightarrow\infty
                  \;,\cr
                          \end{aligned} \right.
\end{equation}
Here  ${B}_\ell $ is  the transmission amplitude. The energy-level shift
$\delta{E}^{B}_{-}$  in two special cases
 can then be calculated.  We find, both when
  $r\rightarrow2M$ and $r\rightarrow\infty$, that
\begin{equation}\label{E-B}
\delta{E}^{B}_{-}=-i\mu^2{\cal{K}}_{B}(-\omega_0)\approx\delta{E}_{0}+\delta{E}_{0r}\;.
\end{equation}
Here, we have defined
\begin{eqnarray}\label{E0}
\delta{E}_{0}=&&-\frac{\mu^2P}{2\pi^2}\int_{0}^{\infty}\frac{\lambda}{\lambda+\omega_0}d\lambda\;,
\end{eqnarray}

\begin{equation}\label{E0r}
\delta{E}_{0r}=-\frac{\mu^2P}{2\pi^2} \int_{0}^{\infty} f(\lambda,r)
\frac{\lambda}{\lambda+\omega_0}d\lambda\;,
\end{equation}
where
\begin{equation}
f(\lambda,r)=\sum_{\ell=0}^\infty
\frac{(1+2l)|\,B_\ell({\lambda}\sqrt{g_{00}})\,|^2}{4 \lambda^2
r^2}\;,
\end{equation}
and $P$ denotes the principal value.
 For the purpose of estimating $\delta{E}^{B}_{-}$ approximately,
we use the geometrical optics approximation to evaluate the
transmission amplitude $B_\ell(p)$~\cite{curvedspace}. In this
approximation,  it is easy to verify that, if $Mp\gg1$, then
transmission ceases when $l$ exceeds $\sqrt{27}Mp$ and this holds
even when $Mp$ is small~\cite{curvedspace}. So we have
\begin{equation}
B_\ell(p)\sim\theta(\sqrt{27}Mp-l)\;,\;\sum_\ell(1+2l)|B_\ell(p)|^2\approx27M^2p^2\;.
\end{equation}
This leads to
\begin{equation}\label{f-lambda-r}
f(\lambda,r)\approx \frac{27M^2g_{00}}{4 r^2}\;.
\end{equation}
 Therefore, we can rewrite  $\delta{E}^{B}_{-}$  in both two special cases as
 \begin{eqnarray}
\delta{E}^{B}_{-}\approx&&\delta{E}_{0}+\frac{27M^2g_{00}}{4
r^2}\delta{E}_{0}\;.
\end{eqnarray}
Here, $\delta{E}_{0}$ is just the shift of the ground state energy
level in the Minkowski vacuum\cite{Audretsch95}. So,
$\delta{E}^{B}_{-}$ contains two divergent terms, both of which
contain a
 linearly divergent factor $\delta{E}_{0}$. According to the Bethe's mass renormalization method\cite{bethe},
 $\delta{E}_{0}$ can be written in two parts
 \begin{equation}
 \delta{E}_{0}=\delta{E'}_{0}+\delta{E''}_{0}=\frac{-\mu^2P}{2\pi^2}\int_{0}^{\infty}d\lambda+\frac{\mu^2P}{2\pi^2}\int_{0}^{\infty}\frac{\omega_0}{\lambda+\omega_0}d\lambda\;,
 \end{equation}
where $\delta{E'}_{0}$ is the energy of a free electron due to its
coupling to the field, which  can also be interpreted as the element
stemming from the renormalization of the mass in the kinetic energy
of the system Hamiltonian. However, it will not contribute to the
observed level shift. The second part
\begin{equation}
\delta{E''}_{0}=\frac{\mu^2P}{2\pi^2}\int_{0}^{\infty}\frac{\omega_0}{\lambda+\omega_0}d\lambda\;,
 \end{equation}
 although  logarithmically divergent,
will give an observable  contribution to the atomic level shift
after a regularization by taking a cutoff of the upper limit of
integration. For the regularization, we assume that the Bethe
method~\cite{bethe} which was first used in a flat spacetime can
also be applied in the present case.  According to Bethe, this
cutoff should be taken as the electron mass $m\footnote{It is worth
pointing out that, in  flat spacetime, the same result as Bethe's
can be obtained if one employs a fully relativistic quantum field
theoretic approach where no cutoff is present. See Refs.
\cite{milonni,Greiner,Kroll-Lamb49,French-Weisskopf49} }.$
Therefore, the renormalized energy-level shift $\delta{E}^{B}_{-}$
reads
\begin{equation}
\delta{E}^{B}_{-}\approx\frac{\mu^2\omega_0}{2\pi^2}\ln\Big(\frac{m}{\omega_0}\Big)\big[1+\frac{27M^2g_{00}}{4
r^2}\big]
\end{equation}
It is easy to see that the regular finite  energy-level shift is
position-dependent, and so generates a force on the atom besides
the gravitational force. This is a force which has a quantum origin
and it is in fact a result of the modified vacuum fluctuations due
to the spacetime curvature. This position-dependent energy shift
gives rise to a force on the atom which can be calculated by taking
the first derivative with respect to $r$. Therefore, we can obtain
the Casimir-Polder-like force on an atom  outside a spherical star
in  two asymptotic regions (i.e., $r\rightarrow\infty $ and
$r\rightarrow2M$):
\begin{equation}\label{F-boulware}
F^B=-\frac{\partial{(\delta{E}^{B}_{-})}}{\partial{r}}\approx\frac{27\mu^2M^2\omega_0}{4\pi^2r^4}(r-3M)\ln\Big(\frac{m}{\omega_0}\Big)
\end{equation}
 A physical realization of the Boulware vacuum would be the vacuum
 state outside a massive spherical body of radius only slightly
 larger than its Schwarzschild  radius.
 According to Eq.~(\ref{F-boulware}), the Casimir-Polder-like force
outside a massive spherical star can be either attractive or
repulsive. In fact, the force is attractive close to the horizon and
repulsive at the spatial asymptotic region with a behavior of
$1/r^3$. The turning point happens near $r\approx 3M$ where the
vacuum field modes are scattered the most. It is interesting to note
that $r=3M$ is the location of the unstable circular orbit of
photons

 \section{ Casimir-Polder-like force for Unruh vacuum }

For a static atom outside a black hole in interaction with
fluctuating massless scalar fields in the Unruh vacuum, let us note
that the field correlation function is given
by~\cite{green1,wightman3}
 \begin{equation}
{G_{U}}^+(x,x')=
\sum_{ml}\int^{\infty}_{-\infty}\frac{e^{-i\omega\Delta{t}}}{4
\pi\omega}|\,Y_{lm}(\theta,\phi)\,|^2\bigg[\frac{|\,\overrightarrow{R_l}(\omega,r)\,|^2}{1-e^{-2\pi\omega/\kappa}}
+\theta(\omega)|\,\overleftarrow{R_l}(\omega,r)\,|^2\bigg]d\omega\;,
 \end{equation}
 where $\kappa=1/(4M)$ is the surface gravity of the black hole.
The corresponding Fourier transform  with  the proper time $\tau$ of
the two-level system  reads
\begin{eqnarray}\label{Guf}
{\cal {G}}_U(\lambda)&=&\int^{\infty}_{-\infty}e^{i{\lambda}\tau}{G_{U}}^+(x,x')d\tau\nonumber\\
&=&\frac{1}{8\pi{\lambda}}\sum_{l=0}^{\infty}\bigg[\theta({\lambda}\sqrt{g_{00}})(1+2l)|\,\overleftarrow{R_l}({\lambda}\sqrt{g_{00}},r)\,|^2
+\frac{(1+2l)|\,\overrightarrow{R_l}({\lambda}\sqrt{g_{00}},r)\,|^2}{1-e^{-2\pi{\lambda}\sqrt{g_{00}}/\kappa}}\bigg]\;.
\end{eqnarray}
A similar calculation as that in the Boulware vacuum yields the
radiative energy-level shift of the ground-state in two special
cases:
\begin{eqnarray}\label{eushift}
\delta{E}^{U}_{-}=-i\mu^2{\cal{K}}_{U}(-\omega_0)\approx\left\{
\begin{aligned}
  &\delta{E}_{0}+\delta{E}_T
+\delta{E}_{0r}\;,&(r\sim{2M})\;;
\\
&\delta{E}_{0}+\delta{E}_{Tr}
+\delta{E}_{0r}\;,&(r\rightarrow\infty)\;.
\end{aligned} \right.
\end{eqnarray}
%%%%%%%%%%%%%%%%%%%%%%%%%%%%%%%%%%%%%
where
\begin{eqnarray}\label{ET}
\delta{E}_T&=& -\frac{\mu^2P}{2\pi^2}\int_{0}^{\infty}
\frac{\lambda}{1-e^{\lambda/T}}\big(\frac{1}{\lambda-\omega_0}
-\frac{1}{\lambda+\omega_0}\big)d\lambda\;,
\end{eqnarray}
and
\begin{eqnarray}\label{ETr}
\delta{E}_{Tr} &=&-\frac{\mu^2P}{2\pi^2} \int_{0}^{\infty}
\frac{f(\lambda,r)\,}{1-e^{\lambda/T}}
\big(\frac{\lambda}{\lambda-\omega_0}-\frac{\lambda}{\lambda+\omega_0}\big)d\lambda\;.
\end{eqnarray}
Here,  $T$  is given by
\begin{equation}
T=\frac{\kappa_r}{2\pi}=\frac{\kappa}{2\pi\sqrt{g_{00}}}=\frac{T_H}{\sqrt{g_{00}}}\;,
\end{equation}
with $T_H=\kappa/2\pi$, being the usual Hawking temperature of the
black hole. This equation is the well-known Tolman
relation~\cite{Tolman304,Tolman301} which gives an effective
temperature as measured by a local observer. A few comments are in
order now. First,  $\delta{E}_T$, which is a contribution to the
energy-level shift of the atom close to the horizon, is
structurally similar to the energy-level shift in a thermal heat
bath at temperature $T$,  while  $\delta{E}_{Tr}$, which is a
contribution to that at infinity, shows the effect of backscattering
of the thermal radiation  off the spacetime curvature, represented
by the grey-body factor $f(\lambda,r)$. These  two terms clearly
support the notion that there is a thermal radiation flux emanating
from the black hole event horizon. Second, the other two terms,
(i.e., $\delta{E}_{0}$ and $\delta{E}_{0r}$) are just terms contained
in $\delta{E}^{B}_{-}$.

%%%%%%%%%%%%%%%%%%%%%%%%%%%%%%%%%%%%%%%%%%%%%%%%%%%%%%%%%
 To analyze the force on the atom in more detail, let us note that
$\delta{E}_T$ can be approximated in the low- and high-temperature
limits as follows
\begin{eqnarray}\label{sh2}
\delta{E}_T&=&-\frac{\mu^2P}{2\pi^2}\int_{0}^{\infty}
\frac{\lambda}{1-e^{\lambda/T}}\big(\frac{1}{\lambda-\omega_0}
-\frac{1}{\lambda+\omega_0}\big)d\lambda
\nonumber\\
&\approx& \left\{
\begin{aligned}
&-\frac{\mu^2T^2}{6\omega_0}-\frac{\mu^2\pi^2T^4}{15\omega_0^3}\;,&
(\omega_0\gg{T})\;;
\\&\frac{\mu^2\omega_0\ln(\omega_0^2/T^2)}{2\pi^2}\;, &(\omega_0\ll{T})\;.
\end{aligned}\right.
\end{eqnarray}
%%%%%%%%%%%%%%%%%%%%%%%%%%%%%%%%%%%
According the  Eq.~(\ref{f-lambda-r}),
 we can  estimate $\delta{E}_{Tr}$ in the same limits
\begin{equation}
\delta{E}_{Tr}=\frac{27M^2g_{00}}{4r^2}\delta{E}_T\approx\left\{
\begin{aligned}
&-\frac{9\mu^2M^2{T}^2g_{00}}{8\omega_0r^2}-\frac{9\mu^2\pi^2M^2{T}^4g_{00}}{20\omega_0^3r^2}\;,&
(\omega_0\gg{T})\;;
\\&\frac{27\mu^2M^2\omega_0 g_{00}}{8\pi^2r^2}\ln\Big(\frac{\omega_0^2}{T^2}\Big)\;,
&(\omega_0\ll{T})\;.
\end{aligned}\right.
\end{equation}
Since close to the horizon, $g_{00}$ approaches zero,
$T=T_H/\sqrt{g_{00}}\gg\omega_0\;$ is always satisfied.
Consequently, the position-dependent energy-level shift can be
written as
\begin{eqnarray}\label{EU-1}
\delta{E^U_{-}}\approx
\frac{\mu^2\omega_0}{2\pi^2}\ln\Big[g_{00}\frac{\omega_0^2}{T_H^2}\Big]+\frac{\mu^2\omega_0}{2\pi^2}\ln\Big(\frac{m}{\omega_0}\Big)\Big[1+\frac{27M^2g_{00}}{4
r^2}\Big]\;.
\end{eqnarray}
When the atom is held static in the spatial asymptotic region, i.e.,
when $r\rightarrow\infty,$  $T\sim{T_H}\;$. Now, the energy-level
shift becomes
\begin{eqnarray}\label{EU-2}
\delta{E^U_{-}}\approx \left\{\begin{aligned}
&-\frac{9\mu^2M^2T_H^2}{8\omega_0 }{1\over
r^2}+\frac{\mu^2\omega_0}{2\pi^2}\ln\Big(\frac{m}{\omega_0}\Big)\Big[1+\frac{27M^2g_{00}}{4
r^2}\Big]\;,~~&(\omega_0\gg{T_H})\;;
\\~
\\&\frac{27\mu^2M^2\omega_0}{8\pi^2r^2}g_{00}\ln\Big[g_{00}\frac{\omega_0^2}{T_H^2}\Big]
+\frac{\mu^2\omega_0}{2\pi^2}\ln\Big(\frac{m}{\omega_0}\Big)\Big[1+\frac{27M^2g_{00}}{4
r^2}\Big]\;,&(\omega_0\ll{T_H})\;.
\end{aligned} \right.
\end{eqnarray}
This position-dependent energy shift gives rise to a force on the
atom which can be calculated by taking the first derivative with
respect to $r$. Close to the horizon, we find that
\begin{equation}\label{F-U-small}
F^U=-\frac{\partial{(\delta{E}^U_{-})}}{\partial{r}}\approx
-\frac{\mu^2M\omega_0}{\pi^2(r-2M)r}-\frac{27\mu^2\omega_0}{64M\pi^2}\ln\Big(\frac{m}{\omega_0}\Big)\;.
\end{equation}
So the Casimir-Polder-like force is attractive and  actually
diverges  at the event horizon. Let us note that the classical force
that is needed to hold the atom static at the horizon also diverges.
If the atom is in the spatial asymptotic region, this force can be
approximated as
\begin{equation}\label{F-U-large}
F^U=-\frac{\partial{(\delta{E}^U_{-})}}{\partial{r}}\approx\left\{
\begin{aligned}
  &-\frac{9\mu^2M^2{T_H}^2}{4\omega_0}{1\over r^3}+\frac{27\mu^2M^2\omega_0}{4\pi^2}\frac{1}{r^3}\ln\Big(\frac{m}{\omega_0}\Big)\;,&(\omega_0\gg{T_H})\;;
\\
&~
\\&-\frac{27\mu^2M^2\omega_0}{2\pi^2}{1\over r^3}\ln\Big(\frac{T_H}{\omega_0}\Big)
+\frac{27\mu^2M^2\omega_0}{4\pi^2}\frac{1}{r^3}\ln\Big(\frac{m}{\omega_0}\Big)\;,&(\omega_0\ll{T_H})\;.
\end{aligned} \right.
\end{equation}
Typically, one has $m\gg\omega_0$, so, far from the black hole, the
force will be attractive if  $T_H\gg{m}$, and repulsive otherwise.
It is interesting to note that the contribution to the
force due to the presence of the Hawking radiation is always
attractive. Therefore, the collapsing of a massive star into a black
hole and the thermal radiation generated as a result makes the
Casimir-Polder-like force more attractive than repulsive.

\section{Conclusion}

In summary, we have calculated the energy-level shift of a two-level
atom outside a spherically symmetric  black hole in the paradigm  of
open quantum systems by looking at the time evolution of the atom
interacting with massless scalar fields in the Boulware vacuum and
the Unruh vacuum. The time evolution of the atom is governed by a
master equation obtained by tracing over the field degrees of
freedom from the complete system. Our results show that, for an atom
in the ground state, the level shift is position-dependent  and
gives rise to a force on the atom besides the classical
gravitational force.

For the case of the Boulware vacuum, which represents  a star that
has not collapsed through its event horizon, this
 force  is attractive  near the horizon, and is repulsive far away from the black hole with
a  behavior of $r^{-3}$ . The  turning point occurs near $r\sim 3M$
where the vacuum field modes are scattered the most.  For the case
of the Unruh vacuum which represents a radiating black hole, we find
that the contribution to the Casimir-Polder-like force due to the
presence of  Hawking radiation is always attractive, and in fact
this attractive force diverges as the horizon is approached.

\begin{acknowledgments}
 One of us (HY) would like to thank the Kavli Institute for Theoretical Physics China where part of this work was done.
 This work was supported in part by the National Natural Science
Foundation of China under Grants No. 11075083, No.11005038 and No.
10935013; the Zhejiang Provincial Natural Science Foundation of
China under Grant No. Z6100077;  the National Basic Research Program
of China under Grant No. 2010CB832803; the PCSIRT under Grant No.
IRT0964;  the Hunan Provincial Natural Science Foundation of China
under Grant No. 11JJ7001; and the Program for the Key Discipline in
Hunan Province.
\end{acknowledgments}

\end{document}